\documentstyle{prhep97}
\makeatletter
\let\chapter\hid@chapter
\makeatother

\newcommand{\bfig}{\begin{center}\begin{picture}}
\newcommand{\efig}[1]{\end{picture}\\{\small #1}\end{center}}

\newcommand{\bmip}[2]{\begin{minipage}[t]{#1pt}\bfig(#1,#2)}
\newcommand{\emip}[1]{\efig{#1}\end{minipage}}

\newcommand{\bq}{\begin{equation}}
\newcommand{\eq}{\end{equation}}
\newcommand{\bqa}{\begin{eqnarray}}
\newcommand{\eqa}{\end{eqnarray}}
\newcommand{\nl}{\nonumber \\}

\begin{document}
\thispagestyle{empty}
\begin{flushright}
CERN-TH/97-327\\
\end{flushright}

\vspace*{3cm}
\begin{center}
\LARGE
{\bf Four-fermion production at LEP2 and NLC}
\normalsize
\footnote{
Contribution to the proceedings of 
{\em The International Europhysics Conference
on High-Energy Physics}, 19-26 August 1997, Jerusalem, Israel.}

\vspace{2cm} 

{\Large ROBERTO PITTAU} \\
\vspace{0.3cm}
{\em Theoretical Physics Division, CERN \\
CH - 1211 Geneva 23, Switzerland}

\vspace*{2cm}  
{\bf ABSTRACT} \\ 
\end{center}
\noindent
The present knowledge on four-fermion
production in electron-positron collisions is reviewed, with 
emphasis on $W$ boson physics. Different methods to extract $M_W$ from
the data are presented and the role of QCD loop corrections discussed.

\vspace*{3.7cm}

\begin{flushleft}
CERN-TH/97-327 \\
November 1997
\end{flushleft}

\clearpage
\setcounter{page}{1}
\authorrunning{R.\,Pittau}
\titlerunning{{\talknumber}: Four-fermion production at LEP2 and NLC}
\def\talknumber{801} 
\title{{\talknumber}: Four-fermion production at LEP2 and NLC}
\author{Roberto\,Pittau
(Roberto.Pittau@cern.ch)}
\institute{Theoretical Physics Division, CERN  CH-1211 Geneva 23}

\maketitle

\begin{abstract}
The present knowledge on four-fermion
production in electron-positron collisions is reviewed, with 
emphasis on $W$ boson physics. Different methods to extract $M_W$ from
the data are presented and the role of QCD loop corrections discussed.
\end{abstract}

\section{Introduction}
Four-fermion processes represent the experimentally measured signal
for $W$ boson physics at LEP2 and NLC.
In fact, the produced $W$'s always decay, giving
four-fermion final states \cite{ref1}.
One is primarily interested in measuring the $W$ mass \cite{ref2}, but also
measurements of the trilinear gauge boson couplings \cite{ref3} and
the four-fermion cross sections \cite{ref4} provide useful information.
An accurate determination of $M_W$ can be combined with the precision
data coming from LEP1 to further constrain the 
Standard Model of the Electroweak Interactions.
In fact, a global fit to the LEP1 data \cite{ref5} predicts the
$W$ mass with an error of the same order as 
the expected final LEP2 error (35 MeV), while NLC will
presumably reduce that error down to 15 MeV \cite{ref6}.
A strong discrepancy between predicted and measured $M_W$ 
would be a signal for new physics. Alternatively, an improvement on
the measurement of $M_W$ can significantly tighten
the present bounds on the Higgs mass through the relation 
\begin{equation}
G_\mu= \frac{\alpha \pi}{\sqrt{2} M^2_W (1-M^2_W/M^2_Z)} \cdot
   \frac{1}{1-\Delta_r(m_t,m_H)}\,, 
\end{equation}
where $\Delta_r$ is a calculable contribution
coming from radiative corrections.

Several tree-level four-fermion codes are available \cite{ref4}.
Electroweak radiative corrections are usually included 
at the leading log (LL) level, 
but very recently, non-factorizable QED corrections have been computed
as well \cite{ref7}. 

\noindent While tree level programs + LL  corrections seem in general 
to be adequate to deal with LEP2 physics, 
further refinements, especially in the sector of the loop radiative 
corrections, are needed in view of the NLC precision physics.

In this contribution, I focus my attention on two 
particular aspects of 
four-fermion $W$ physics, namely $M_W$ measurement and
QCD corrections. 

In the next section, I describe two techniques to measure
the $W$ mass, and present a new method to extract $M_W$ from
the best measured variables.

\noindent When quarks are present in the final state, 
QCD loop contributions have to be included as well. 
Those corrections are discussed in the last section of the paper. 

\section{$M_W$ measurement}
Two methods are mainly used to extract the $W$ mass:
the threshold method and the direct reconstruction technique \cite{ref2}.
In the first case the total $W^+ W^-$ cross section is measured near
threshold (161 GeV), where the sensitivity to $M_W$ is stronger, and
plotted as a function of the $W$ mass.
At LEP2, that gives $M_W= 80.40 \pm 0.22$ GeV \cite{ref5}.

The direct reconstruction method is applied at higher energy,
where the statistics increases. It requires three steps:
\begin{enumerate}
\item  From the experimental data the invariant mass distribution
$\frac{d\sigma}{dM}$ is reconstructed. To improve the mass 
resolution, a constrained fit is usually performed event by
event, assuming no Initial State Radiation (ISR) and equality between  
the invariant masses coming from different $W$'s.
\item  A theoretical distribution is taken for
$\frac{d\sigma}{dM}$ (usually a convolution of a Breit-Wigner with
a Gaussian) and a mass $M_W^\prime$ fitted.
\item  The value  $M_W^\prime$ is then corrected by
Monte Carlo, from the bias introduced by the constrained fit, to
get the reconstructed $W$ mass $M_R$ with an error $\Delta M_R$ .
\end{enumerate}
Such a measurement gave $M_W= 80.37 \pm 0.19$ GeV \cite{ref5}, in the LEP2
run at 172 GeV. 

Recently, a new method has been proposed \cite{ref8}
(direct fit method), in which only the best 
measured quantities are used to extract the $W$ mass. The idea is simple.
Given a set of well measured quantities $\{\Phi\}$ one 
computes, event by event, the theoretical probability $P_i$ of
getting the observed set of values $\{\Phi_i\}$ for $\{\Phi\}$. 
This is a function of $M_W$ and is given
by the ratio of the
differential cross section in those variables, divided by the
total cross section in the experimental fiducial volume
\begin{equation}
  P_i(M_W)= \frac{\frac{d\sigma}{d \Phi_i}}{\sigma}\,.         
\end{equation}
Given $N$ observed events, 
the logarithm of the likelihood function $L$ \begin{equation}
\log\,L(M_W) \equiv \log\,\prod_{i=1}^N P_i(M_W)=
              \sum_{i=1}^N \log\frac{d\sigma}{d \Phi_i}(M_W)
             -N \log \sigma(M_W)
\end{equation}
is distributed, for large $N$, as a quadratic function of $M_W$.
The previous equation is then computed for different values of
$M_W$ and a parabola fitted, from which the reconstructed $W$ mass
$M_R$ is obtained with an error $\Delta M_R$.

\noindent In order to construct a tool for the evaluation of $P_i(M_W)$
one has to choose the set $\{\phi\}$ of accurately measured variables.
Although one can always consider more sets $\{\phi\}$,  the
following choices seem reasonable in practice \cite{ref9} for different 
four-fermion final states:

\begin{enumerate}
\item Semileptonic case: $q_1q_2\ell\nu$ \\
1a $\{\phi\} = \{E_\ell, \Omega_\ell, \Omega_{q_1}, \Omega_{q_2}\}$ \\
1b $\{\phi\} = \{E_\ell, \Omega_\ell, \Omega_{q_1}, \Omega_{q_2}, E_h\}$, where
$E_h$ is the total energy of the jets.
\item Purely hadronic case: $q_1q_2q_3q_4$ \\
$\{\phi\} =\{\Omega_{q_1}\Omega_{q_2}\Omega_{q_3}\Omega_{q_4}\}$.
\item Purely leptonic case: $\ell_1 \nu_1 \ell_2 \nu_2$\\
$\{\phi\} = \{E_{\ell_1}, \Omega_{\ell_1}, E_{\ell_2},\Omega_{\ell_2}\}$
\end{enumerate}

\noindent Since eight variables determine an event 
when no ISR is present, sets 1a
and 3 would require one or two integrations, cases 1b  and 2 none. 
Including ISR adds two integrations. When jets cannot be assigned to
specific quarks a folding over the various possibilities should be
included.

As an example of the direct fit method,
I show, in table 1, the reconstructed masses obtained by fitting a sample
of 1600 EXCALIBUR \cite{ref1} CC3 events \cite{ref4} including ISR,
generated with an input mass of $M_W= 80.23$ GeV, at $\sqrt{s}=$ 190 GeV.
In all four cases, the $W$ mass is correctly reconstructed.

\begin{table}[h]
\begin{center}
\begin{tabular}{|c|c|c|c|} \hline 
1a               & 1b       & 2        & 3 \\ \hline\hline
 \rule[-7 pt]{0 pt}{24 pt}
80.238 $\pm$ 0.049&80.238 $\pm$ 0.032 &80.255 $\pm$ 0.036&80.209 $\pm$
 0.077\\ 
\hline
\end{tabular}
\end{center}
\caption{Reconstructed $W$ mass (GeV) with four choices of the set
$\{\Phi\}$ (see text).}
\end{table}
Cases 1b and 2 give better errors, because less information
is integrated
out. Conversely, in the leptonic case 3, the error is worse, 
since a large part of the kinematical information is actually
missing.

In the direct reconstruction method, one wants to keep
as much information as possible, also preferably photon
momenta, in order to reconstruct the kinematics event by event.
On the contrary, in the direct fit method, one has to integrate over all
information that is not well determined, in particular ISR.
Because of the fact that the integral over the $p_T$ distribution
of ISR photons is theoretically better known than the distribution itself,
one expects the details of the radiation to matter less in the direct
fit method.
That can be viewed as an advantage with respect to the 
direct reconstruction technique.

However, a last remark is in order. All numbers presented in table 1
refer to the partonic level, without inclusion of hadronization
effects and detector resolution. Therefore one still
has to prove that the fitting procedure survives those effects.
This question is currently under investigation \cite{ref10}.

It is also clear that the direct fit method is not only applicable to
measure $M_W$, but can be used, in principle, to extract any
parameter - as $\Gamma_W$ or a set of anomalous trilinear gauge 
couplings (TGCs) - from the data sample. Not surprisingly, 
the whole strategy for the direct fit method has been first discussed 
in the final report of the Workshop on Physics at LEP2, 
in the context of TGCs determination \cite{ref3}.   

\section{QCD corrections}
QCD loop corrections to four-fermion production in $e^+ e^-$ collisions can
be divided in two classes, namely QCD corrections 
to ${\cal O}(\alpha^2 \alpha_s^2)$ and ${\cal O} (\alpha^4)$ processes,
respectively.

The first corrections appear as ${\cal O} (\alpha_s^2)$ contributions
to four-jet production via QCD \cite{ref11}, 
while the second ones are relevant 
for studying $W$ boson physics, and, more in general, semileptonic four-fermion
processes and fully hadronic final states mediated by electroweak
bosons. 

I shall concentrate here on the latter contributions,
which can all be obtained by defining suitable combinations of loop 
diagrams plus real gluon radiation, as shown in ref. \cite{ref12}. 
\noindent The calculation is simplified a lot by using the
reduction procedure presented in ref. \cite{ref13}.

In table 2, cross sections computed with the program in
ref. \cite{ref12} are presented, for the 
semileptonic process $e^+ e^- \to \mu^- \bar \nu_\mu u \,\bar d$. 

\begin{table}[h]
\begin{center}
\begin{tabular}{|c|c|c|c|} \hline 
$\sqrt{s}$ & Born & NLO & nQCD \\ \hline\hline
 \rule[-7 pt]{0 pt}{24 pt}
$161$ GeV &.24962 $\pm$ .00002 &.24760 $\pm$ .00002 &.24790 $\pm$ .00002\\ 
\hline
 \rule[-7 pt]{0 pt}{24 pt}
$175$ GeV &.96006 $\pm$ .00007 &.94519 $\pm$ .00007 &.94613 $\pm$ .00007\\ 
\hline
 \rule[-7 pt]{0 pt}{24 pt}
$190$ GeV &1.184003 $\pm$ .00009 &1.16681 $\pm$ .00009 &1.16766 $\pm$ .00008\\ 
\hline
 \rule[-7 pt]{0 pt}{24 pt}
$500$ GeV &.46970 $\pm$ .00006 &.47109 $\pm$ .00007&.46131 $\pm$ .00006 \\ 
\hline
\end{tabular}
\end{center}
\caption{Cross sections in pb for $e^+ e^- \to \mu^- \bar \nu_\mu u \,\bar
 d$ with canonical cuts \cite{ref4} .}
\end{table}
\noindent The exact calculation 
(NLO) is compared with a ``naive'' approach to strong radiative corrections
(nQCD), where the QCD contributions are simply included 
through the substitutions
\begin{equation}
\Gamma_W \to \Gamma_W\,\left( 1+\frac{2}{3}\,\frac{\alpha_s}{\pi}
\right)\,,
~~~~~~\sigma \to \sigma\,\left(1+\frac{\alpha_s}{\pi}\right) \,.
\end{equation}
For the direct reconstruction of the $W$ mass, the quantity

\noindent $\langle \Delta M \rangle = \frac{1}{2\sigma}  
     \int~(\sqrt{s_+}+\sqrt{s_-}-2\,M_W)~d\sigma$ is relevant \cite{ref4},
where $M_W$ is the input mass in the program.
One gets, with canonical cuts at $\sqrt{s}= 175$ GeV:
\bqa
\langle \Delta M \rangle_{{\rm NLO}} = &-& 0.5585 \pm 0.0002~{\rm GeV} \nl
\langle \Delta M \rangle_{{\rm nQCD}}= &-& 0.5583 \pm 0.0002~{\rm GeV}\,. 
\eqa
Also, one can show that the angular distributions are distorted, 
in the NLO calculation, with
respect to the nQCD prediction \cite{ref12}.

  From the previous results it is clear that 
nQCD is adequate at LEP2 for semileptonic processes,
but exact calculations are important at NLC and for anomalous
couplings studies, where
the angular distributions matter to constrain the anomalous contributions.

In table 3, I show results for the fully hadronic process
$e^+ e^- \to u \bar d s \bar c$ at $\sqrt{s}=$ 175 GeV.
The numbers are obtained by using the program described in ref. \cite{ref14}.
In case (a) only canonical cuts are applied.
In (b) two reconstructed masses $M_{R1}$ and $M_{R2}$ are determined by 
minimizing the quantity $ \Delta_M^\prime= (M_{R1}-M_W)^2+
                                           (M_{R2}-M_W)^2$,
and a cut $| M_{Ri}-M_W | < 10$ GeV is imposed.
In (c) a smearing with a Gaussian with a 2 GeV width
is introduced in addition, to mimic the experimental resolution.

\begin{table}[h]
\begin{center}
\begin{tabular}{|c|c|c|c|} \hline 
$\sigma({\rm pb})$             & (a)& (b) & (c)\\
\hline\hline
 \rule[-7 pt]{0 pt}{24 pt}
${\rm NLO}$ &1.1493(4)      &0.7895(5)      &0.7758(9) \\ 
\hline
 \rule[-7 pt]{0 pt}{24 pt}
${\rm nQCD}$ &1.1069(3)      &1.0545(3)      &1.0479(3) \\ 
\hline
\end{tabular}
\end{center}
\caption{Cross sections for $e^+ e^- \to u \bar d s \bar c$
at $\sqrt{s}=$ 175 GeV.}
\end{table}
\noindent For the process at hand one gets, for case (b),
\bqa
\langle \Delta M \rangle_{{\rm NLO}} = &-& 0.2290 \pm 0.0010~{\rm GeV} \nl
\langle \Delta M \rangle_{{\rm nQCD}}= &-& 0.0635 \pm 0.0004~{\rm GeV}\,,
\eqa
where now $\langle\Delta M\rangle = \frac{1}{2\sigma}  
     \int~(M_{R1}+M_{R2}-2\,M_W)~d\sigma$.

We are easily convinced, by the above results, 
that the naive QCD implementation fails in describing hadronic
four-fermion final states at LEP2. In particular, 
the reduction in cross section 
(compare cases (a) and (b) in table 3) shows that many
soft gluons are exchanged between decay products of different $W$'s
that are not taken into account using nQCD.
A failure of nQCD can also be proved at NLC energies \cite{ref14}.
\section{Conclusions}
Precise measurements of $M_W$ are performed at LEP2, using the
threshold method and the direct reconstruction of 
$\frac{d\sigma}{dM_W}$. 

An alternative technique is also available in which only well
measured quantities are directly used to extract $M_W$ (and 
TGCs) from the data.

QCD loop corrections to semileptonic
${\cal O}(\alpha^4)$ processes are well approximated at LEP2, by nQCD,
except for angular distributions, while the naive approach fails in describing
hadronic four-fermion final states.

All existing calculations have to be refined and radiative
corrections better understood in view of the NLC precision measurements.

\end{document}